\documentclass[aps,prl,groupedaddress,preprint,showpacs]{revtex4}

\usepackage{graphicx}
\usepackage{amssymb}

\def\be{\begin{equation}}
\def\ee{\end{equation}}
\def\ba{\begin{eqnarray}}
\def\ea{\end{eqnarray}}
\def\bc{\begin{center}}
\def\ec{\end{center}}

\begin{document}

\title{Revision of the theory of a uniform electron gas
}

\author{S. A. Mikhailov
}
\affiliation{Institute of Physics, University of Augsburg, D-86135 Augsburg, Germany}

\date{\today}
\begin{abstract}
The many-body theory of a uniform electron gas was developed at the end of 1950ies. The Coulomb interaction between electrons was considered as a perturbation and the ground state energy was calculated in all orders of the non-degenerate perturbation theory. We show that the ground state of the unperturbed system is actually strongly degenerate and that the degenerate perturbation theory leads to lower values of the ground state energy.
\end{abstract}


 
\maketitle

The model of a uniform electron gas is very important in the many-body theory. It considers a system of Coulomb interacting electrons immersed in a uniformly distributed positive background (jellium). Although all effects related to the discrete distribution and the motion of atoms in the crystal lattice of a real solid are completely ignored, the electron-electron interaction is supposed to be treated exactly in this model. Therefore it provides a valuable step towards  understanding of the electron-electron interaction phenomena in solids. The theory of the uniform electron gas was worked out by many authors more than 50 years ago \cite{GellMann57,Sawada57,Hubbard57,Nozieres58,Quinn58} and nowadays it can be found in many textbooks on the condensed matter theory, see e.g. Refs. \cite{Mahan90,Kittel87,Fulde95}.

In the jellium model one considers $N$ interacting electrons in the volume $\Omega$. The average density of the electrons $n_v=N/\Omega$ is assumed to be equal to the density of the uniform positive background to ensure electroneutrality. The Hamiltonian of the system reads
\be
\hat H= \hat H_0+V=\hat K+V_{ee}+V_{eb}+V_{bb}=\sum_{i=1}^N\frac{{\bf \hat p}_i^2}{2m} + \frac 12 \sum_{i\neq j=1}^N \frac {e^2}{|{\bf r}_i-{\bf r}_j |} +V_{eb}+V_{bb},\label{Hm}
\ee
where $\hat H_0=\hat K$ is the kinetic energy of $N$ electrons and $V=V_{ee}+V_{eb}+V_{bb}$ is the total Coulomb energy of the system which consists of the electron-electron $V_{ee}$, electron-background $V_{eb}$ and background-background $V_{bb}$ interaction energies.  

One of the most important physical quantity calculated by the theory is the ground state energy $E_{g}$ of the Hamiltonian (\ref{Hm}). The result of such calculations \cite{GellMann57}, the energy $E_{g}$ per particle, measured in Rydbergs $R= me^4/2\hbar^2$, 
\be
\frac{E_{g}}{N R}=\frac{2.21}{r_s^2}-\frac{0.916}{r_s}+0.0622\ln(r_s)-0.094+\dots,\label{en}
\ee
is expressed in terms of the dimensionless density parameter 
\be
r_s=\frac{r_0}{a_B}=\frac{me^2}{\hbar^2}\left(\frac 3{4\pi n_v}\right)^{1/3},\label{rs}
\ee
defined as the ratio of the radius of the sphere $r_0$, which encloses one electron, to the Bohr radius $a_B=\hbar^2/me^2$. The $r_s$-parameter is proportional to $e^2$, i.e. the series (\ref{en}) is an expansion in powers of the Coulomb interaction strength, and is inversely proportional to $n_v^{1/3}$, i.e. the small-$r_s$ limit corresponds to the high electron densities. The first term in the $E_g(r_s)$ expansion is $e^2$-independent and corresponds to the kinetic energy of the electrons. The second term is the exchange Coulomb energy. All the rest terms in the expansion (\ref{en}) are referred to as the correlation energy \cite{Mahan90,Kittel87,Fulde95}. The formula (\ref{en}) is widely used in the density functional theory. 

Similar calculations were also done for a two-dimensional (2D) uniform electron gas \cite{Rajagopal77,Isihara77a}. In the 2D case the density parameter $r_s$ is defined as 
\be
r_s=r_0/a_B=1/\sqrt{\pi n_s}a_B,
\ee
where $n_s$ is the surface electron density, and the result reads 
\be
\frac{E_{g}}{N R}=
\frac 1{r_s^2}-\frac {1.2004}{r_s}-0.38-0.172 r_s\ln r_s+\dots 
\label{en2}.
\ee
Equations (\ref{en}) and (\ref{en2}) were obtained by applying the perturbation theory to the many-body Schr\"odinger equation $\hat H\Psi=(\hat H_0+V)\Psi=E\Psi$, where $\hat H_0$ corresponded to the non-perturbed problem and the total Coulomb energy $V$ was treated as a perturbation.

In this Letter we show that the method that lead to the formulas (\ref{en}) and (\ref{en2}) was not fully adequate. Deriving the formulas (\ref{en}) and (\ref{en2}) one assumed that the unperturbed ground state wave function $|\Psi_g^0\rangle$ is given by a single Slater determinant $|\Psi_g^0\rangle \propto \det |\psi_{{\bf k}_i}({\bf r}_j)|$ composed of the plane waves $\psi_{{\bf k}_i}({\bf r}_j)\propto \exp(i{\bf k}_i\cdot {\bf r}_j)$ with the wavevectors ${\bf k}_i$ lying under the Fermi surface $|{\bf k}_i|\le k_F$. The second (exchange) and all subsequent terms in the expansions (\ref{en}) and (\ref{en2}) were obtained as the first- ($\langle \Psi_g^0|V|\Psi_g^0\rangle$) and higher-order corrections  of the {\em non-degenerate} perturbation theory. 

The ground state of the unperturbed Hamiltonian $\hat H_0$ is, however, {\em highly degenerate}. Consider, for example, a 2D electron gas in the shape of an $L\times L$ square and assume for simplicity that we have only $N=4$ electrons. Applying as usual the periodic boundary conditions we get the allowed values of the wavevector ${\bf k}=(2\pi/L)(m,n)$, where $m$ and $n$ are integers. The  wavevectors ${\bf k}_i$ of each of these four electrons should be appropriately chosen to minimize the expectation value of the kinetic energy $\hat K$. The first pair of electrons (with the spins $\sigma$ up and down) should evidently be placed in the state $(m,n)=(0,0)$. The third and fourth electrons have however a broad choice. The next energy ``shell'', with the ``energy'' $m^2+n^2=1$, has four possible $(m,n)$-places, $(m,n)=(1,0)$, $(0,1)$, $(-1,0)$ and $(0,-1)$, which give eight single-particle states (accounting for two spin projections). Two electrons can be distributed over the eight allowed states by ${_8C_2}=8!/2!6!=28$ ways. All these 28 many-body states have the same energy in the zeroth order of the perturbation theory. All many-body states can be classified according to their total momentum ${\bf K}^{tot}= \sum_{i}^N{\bf k}_i$, the total spin $S^{tot}$ and the projection of the total spin $S_z^{tot}$ (the conserving quantum numbers). Among the 28 degenerate lowest-energy states there are the states 
\begin{enumerate}
\item[(a)] with a finite total momentum ${\bf K}^{tot}\neq {\bf 0}$ and the finite total-spin projection $S_z^{tot}=1$ or $-1$, e.g. 
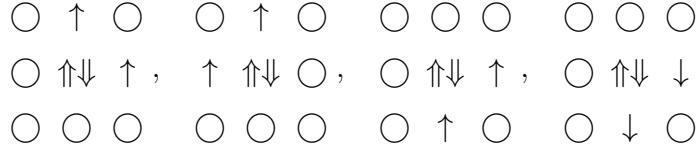
$$
\begin{array}{ccc}
\bigcirc& \uparrow & \bigcirc \\
\bigcirc& \Uparrow\Downarrow & \uparrow \\
\bigcirc& \bigcirc & \bigcirc \\
\end{array}, \ \ 
\begin{array}{ccc}
\bigcirc& \uparrow & \bigcirc \\
\uparrow & \Uparrow\Downarrow & \bigcirc\\
\bigcirc& \bigcirc & \bigcirc \\
\end{array}, \ \ 
\begin{array}{ccc}
\bigcirc& \bigcirc & \bigcirc \\
\bigcirc& \Uparrow\Downarrow & \uparrow \\
\bigcirc& \uparrow & \bigcirc \\
\end{array}, \ \ 
\begin{array}{ccc}
\bigcirc& \bigcirc & \bigcirc \\
\bigcirc& \Uparrow\Downarrow & \downarrow \\
\bigcirc& \downarrow & \bigcirc \\
\end{array}
$$
[shown is the $(k_x,k_y)$-plane with the unoccupied (circles) and occupied (arrows) states; up- and down-arrows symbolize up and down spin-polarized electrons; the symbols $\Uparrow\Downarrow$ in the center of each drawing correspond to the always occupied state $(m,n,\sigma)=(0,0,\uparrow)$ and $(0,0,\downarrow)$], 
\item[(b)] with a finite total momentum ${\bf K}^{tot}\neq {\bf 0}$ and the zero spin projection $S_z^{tot}=0$, e.g. 
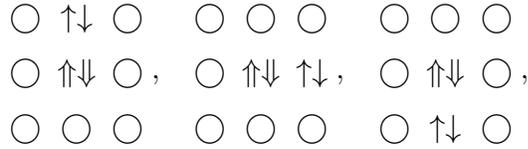
$$
\begin{array}{ccc}
\bigcirc& \uparrow\downarrow & \bigcirc \\
\bigcirc& \Uparrow\Downarrow & \bigcirc \\
\bigcirc& \bigcirc & \bigcirc \\
\end{array}, \ \ 
\begin{array}{ccc}
\bigcirc& \bigcirc& \bigcirc \\
\bigcirc& \Uparrow\Downarrow & \uparrow\downarrow  \\
\bigcirc& \bigcirc & \bigcirc \\
\end{array}, \ \ 
\begin{array}{ccc}
\bigcirc& \bigcirc & \bigcirc \\
\bigcirc& \Uparrow\Downarrow & \bigcirc \\
\bigcirc& \uparrow\downarrow & \bigcirc \\
\end{array},
$$
\item[(c)] with the vanishing total momentum ${\bf K}^{tot}={\bf 0}$ and the finite total spin projection, e.g. 
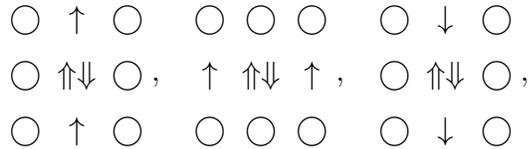
$$
\begin{array}{ccc}
\bigcirc& \uparrow & \bigcirc \\
\bigcirc& \Uparrow\Downarrow & \bigcirc\\
\bigcirc& \uparrow  & \bigcirc \\
\end{array}, \ \ 
\begin{array}{ccc}
\bigcirc&  \bigcirc & \bigcirc \\
\uparrow & \Uparrow\Downarrow & \uparrow\\
\bigcirc& \bigcirc & \bigcirc \\
\end{array}, \ \ 
\begin{array}{ccc}
\bigcirc& \downarrow & \bigcirc \\
\bigcirc& \Uparrow\Downarrow & \bigcirc \\
\bigcirc& \downarrow & \bigcirc \\
\end{array}, 
$$
\item[(d)] and with the zero total momentum ${\bf K}^{tot}={\bf 0}$ and the zero spin projection $S_z^{tot}=0$, e.g. 
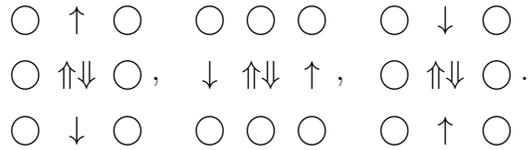
$$
\begin{array}{ccc}
\bigcirc& \uparrow & \bigcirc \\
\bigcirc& \Uparrow\Downarrow & \bigcirc \\
\bigcirc& \downarrow & \bigcirc \\
\end{array}, \ \ 
\begin{array}{ccc}
\bigcirc& \bigcirc& \bigcirc \\
\downarrow & \Uparrow\Downarrow & \uparrow \\
\bigcirc& \bigcirc & \bigcirc \\
\end{array}, \ \ 
\begin{array}{ccc}
\bigcirc& \downarrow & \bigcirc \\
\bigcirc& \Uparrow\Downarrow & \bigcirc \\
\bigcirc& \uparrow & \bigcirc \\
\end{array}.
$$
\end{enumerate}
In general, it is unknown in advance whether the ground state of the Hamiltonian $\hat H$ will have the vanishing total momentum ${\bf K}^{tot}={\bf 0}$ and the vanishing total spin $S^{tot}=0$, as it was actually assumed in (\ref{en}), (\ref{en2}), or it will correspond to a superconducting ground state with ${\bf K}^{tot}\neq {\bf 0}$ and/or to a partly spin-polarized ground state with $S^{tot}\neq 0$. The answer to this question can be obtained only by means of  direct calculations taking into account a sufficiently large number of basis many-body states (Slater determinants). Whether the true ground state has zero or finite ${\bf K}^{tot}$ and $S^{tot}$ may also depend on the Coulomb interaction strength, i.e. on the $r_s$-parameter, cf.  Refs. \cite{Mikhailov02b,Mikhailov02e}.

But assume for the moment that the many-body ground state does correspond to ${\bf K}^{tot}={\bf 0}$ and $S_z^{tot}=S^{tot}=0$. The following four many-body states correspond to ${\bf K}^{tot}={\bf 0}$ and $S_z^{tot}=0$:
$$
|A\rangle\sim \left(\begin{array}{ccc}
\bigcirc& \uparrow & \bigcirc \\
\bigcirc& \Uparrow\Downarrow & \bigcirc \\
\bigcirc& \downarrow & \bigcirc \\
\end{array}\right), \ \ 
|A'\rangle\sim \left(\begin{array}{ccc}
\bigcirc& \downarrow & \bigcirc \\
\bigcirc& \Uparrow\Downarrow & \bigcirc \\
\bigcirc& \uparrow & \bigcirc \\
\end{array}\right), \ \ 
|B\rangle\sim \left(\begin{array}{ccc}
\bigcirc& \bigcirc& \bigcirc \\
\downarrow & \Uparrow\Downarrow & \uparrow \\
\bigcirc& \bigcirc & \bigcirc \\
\end{array}\right), \ \ 
|B'\rangle\sim \left(\begin{array}{ccc}
\bigcirc& \bigcirc & \bigcirc \\
\uparrow & \Uparrow\Downarrow & \downarrow \\
\bigcirc& \bigcirc & \bigcirc \\
\end{array}\right).
$$
The states $(|A\rangle +|A'\rangle)/\sqrt{2}$ and $(|B\rangle +|B'\rangle)/\sqrt{2}$ have the total spin $S^{tot}=1$, while the linear combinations 
\be
\Psi_1\sim \frac {|A\rangle-|A'\rangle}{\sqrt{2}} \sim \frac {1}{\sqrt{2}}
\left[
\left(\begin{array}{ccc}
\bigcirc& \uparrow & \bigcirc \\
\bigcirc& \Uparrow\Downarrow & \bigcirc \\
\bigcirc& \downarrow & \bigcirc \\
\end{array}\right) - \left(\begin{array}{ccc}
\bigcirc& \downarrow & \bigcirc \\
\bigcirc& \Uparrow\Downarrow & \bigcirc \\
\bigcirc& \uparrow & \bigcirc \\
\end{array}\right)
\right]
\ee
and 
\be
\Psi_2\sim \frac {|B\rangle-|B'\rangle}{\sqrt{2}} \sim \frac {1}{\sqrt{2}}
\left[
\left(\begin{array}{ccc}
\bigcirc& \bigcirc& \bigcirc \\
\downarrow & \Uparrow\Downarrow & \uparrow \\
\bigcirc& \bigcirc & \bigcirc \\
\end{array}\right) - 
\left(\begin{array}{ccc}
\bigcirc& \bigcirc & \bigcirc \\
\uparrow & \Uparrow\Downarrow & \downarrow \\
\bigcirc& \bigcirc & \bigcirc \\
\end{array}\right)
\right]
\ee
have the desired properties ${\bf K}^{tot}={\bf 0}$, $S_z^{tot}=S^{tot}=0$. The states $\Psi_1$ and $\Psi_2$ have the same kinetic energy, $\langle \Psi_1 | \hat H_0 | \Psi_1 \rangle = \langle \Psi_2 | \hat H_0 | \Psi_2 \rangle \equiv E_0=2(\hbar^2/2m)(2\pi/L)^2$, i.e. the ground state of the unperturbed problem is doubly degenerate (under the conditions ${\bf K}^{tot}={\bf 0}$, $S_z^{tot}=S^{tot}=0$). The solution of the Schr\"odinger problem should therefore be searched for in the form $\Psi=C_1\Psi_1 + C_2\Psi_2$. The energy levels of the Hamiltonian $\hat H$ are then determined from the equation
\be
\left(
\begin{array}{cc}
E_0+V_{11}-E & V_{12} \\
V_{21} & E_0+V_{22}-E \\
\end{array}
\right) 
\left(
\begin{array}{c}
C_1 \\
C_2 \\
\end{array}
\right) 
=0,\label{4pen}
\ee
where $V_{ij}=\langle\Psi_i|V|\Psi_j\rangle$ and $V_{11}=V_{22}$, $V_{12}=V_{21}$ due to symmetry. As the off-diagonal Coulomb matrix elements are non-zero, in general, thus calculated ground state energy  
\be
E_g=E_0+V_{11}-|V_{12}|
\ee
is evidently lower than the one ($E_0+V_{11}$) one would get using the non-degenerate theory. Since both $V_{11}$ and $V_{12}$ are proportional to the first power of $r_s$, the described procedure {\em reduces the numerical coefficient} in the exchange energy terms [0.916 in (\ref{en}) and 1.200 in (\ref{en2})], i.e. this correction is essential even in the limit $r_s\to 0$.

In the considered simplest case of only four electrons the diagonal Coulomb matrix elements are the same, $V_{11}=V_{22}$, and the difference from the textbook solution (\ref{en}), (\ref{en2}) arises only due to the off-diagonal Coulomb matrix elements. This is not always so. Consider for example another simple case with $N=6$ electrons. Now four electrons should be distributed over the eight quantum states on the last partly-occupied shell. This gives the total degeneracy $D$ of the unperturbed ground state $D={_8C_4}=70$. Restricting ourselves by only the many-body states with ${\bf K}^{tot}={\bf 0}$ and $S_z^{tot}=0$ we have to consider eight degenerate states, namely the states (group A)
$$
|1\rangle\sim \left(\begin{array}{ccc}
\bigcirc& \uparrow & \bigcirc \\
\uparrow& \Uparrow\Downarrow & \downarrow \\
\bigcirc& \downarrow & \bigcirc \\
\end{array}\right), \ \ 
|2\rangle\sim \left(\begin{array}{ccc}
\bigcirc& \uparrow & \bigcirc \\
\downarrow& \Uparrow\Downarrow & \uparrow \\
\bigcirc& \downarrow & \bigcirc \\
\end{array}\right), \ \ 
|3\rangle\sim \left(\begin{array}{ccc}
\bigcirc& \downarrow& \bigcirc \\
\downarrow & \Uparrow\Downarrow & \uparrow \\
\bigcirc& \uparrow & \bigcirc \\
\end{array}\right), \ \ 
|4\rangle\sim \left(\begin{array}{ccc}
\bigcirc& \downarrow & \bigcirc \\
\uparrow & \Uparrow\Downarrow & \downarrow \\
\bigcirc& \uparrow & \bigcirc \\
\end{array}\right),
$$
and the states (group B)
$$
|5\rangle\sim \left(\begin{array}{ccc}
\bigcirc& \uparrow & \bigcirc \\
\downarrow& \Uparrow\Downarrow & \downarrow \\
\bigcirc& \uparrow & \bigcirc \\
\end{array}\right), \ \ 
|6\rangle\sim \left(\begin{array}{ccc}
\bigcirc& \downarrow & \bigcirc \\
\uparrow& \Uparrow\Downarrow & \uparrow \\
\bigcirc& \downarrow & \bigcirc \\
\end{array}\right), \ \ 
|7\rangle\sim \left(\begin{array}{ccc}
\bigcirc& \uparrow\downarrow& \bigcirc \\
\bigcirc & \Uparrow\Downarrow & \bigcirc \\
\bigcirc& \uparrow\downarrow & \bigcirc \\
\end{array}\right), \ \ 
|8\rangle\sim \left(\begin{array}{ccc}
\bigcirc& \bigcirc & \bigcirc \\
\uparrow\downarrow & \Uparrow\Downarrow & \uparrow\downarrow \\
\bigcirc& \bigcirc & \bigcirc \\
\end{array}\right).
$$
The diagonal Coulomb matrix elements $\langle n|V|n\rangle$ are the same for the states of the group A and for the states of the group B, but they are not equal to each other,
\be
V_{11}=V_{22}=V_{33}=V_{44}\neq V_{55}=V_{66}=V_{77}=V_{88}.
\ee
Apart from the non-zero off-diagonal terms the $8\times 8$ Hamiltonian matrix will now have different diagonal terms. 

\begin{figure}
\includegraphics[width=13.5cm]{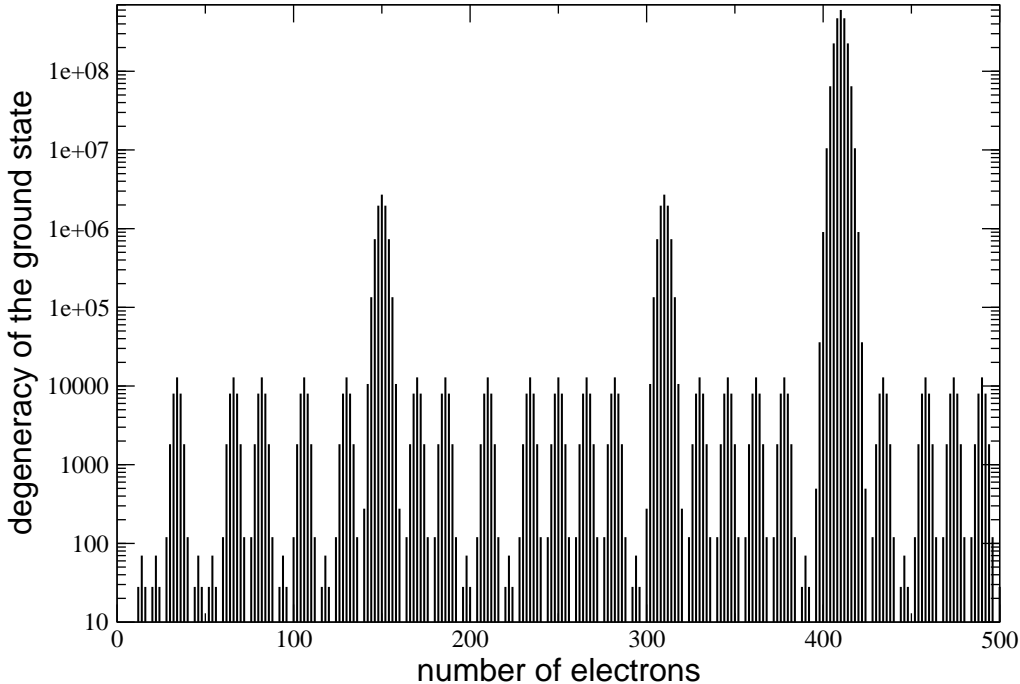}
\caption{The ground state degeneracy of the unperturbed Hamiltonian $H_0$ as a function of the particle number $N$. \label{degen}}
\end{figure}

Obviously, the ground state degeneracy of the unperturbed Hamiltonian $H_0$ dramatically grows with the number of particles $N$, see Figure \ref{degen}. For example, for $N=410$ the full degeneracy (including the states with the finite total momentum and spin) is $D={_{32}C_{16}}=601\,080\,390$ and for $N=2042$ it is $D={_{48}C_{24}}=32\,247\,603\,683\,100$. If to take into account only the states with ${\bf K}^{tot}={\bf 0}$ and $S^{tot}=S_z^{tot}=0$, the degeneracy will be somewhat lower but it will still be an exponentially huge number.

For some numbers $N$ (e.g. for $N=10$, 18, 26, ..., 426, ...), corresponding to the fully occupied shells (analogs of the noble gase atoms), the non-perturbed ground state is indeed non-degenerate ($D=1$). For these numbers the non-degenerate perturbation theory is formally correct. However, taking into account many-body states which include the single-particle states from the next empty shell, one can also reduce $E_g$ as compared to the non-degenerate theory. The loss in the kinetic energy is smaller in this case than the gain in the Coulomb energy obtained due to the larger basis set of many-body functions. For example, for $N=26$ two-dimensional electrons (four fully occupied shells) we found that the ground state energy can be reduced by $\sim 0.32\%$ if to let two electrons occupy the empty fifth shell (these calculations have been done by the method similar to that described in \cite{Mikhailov02b}). 

In order to get a reasonably accurate solution of the problem at finite $r_s$ one had to diagonalize a very large Hamiltonian matrix \cite{GellMann}. The size of the matrix grows with the number of particles $N$, dimensionality of space (2D $\rightarrow$ 3D)  and the density parameter $r_s$. At $r_s\simeq 1$ any practically realistic calculation of this type could be made only for $N\lesssim 10$ (cf. Refs. \cite{Mikhailov02b,Mikhailov02e,Merkt91,Wagner92,Mikhailov02c,Tavernier03}). The impression, given by Eqs. (\ref{en}), (\ref{en2}), that one can exactly solve the many-body Coulomb problem in the thermodynamic limit $N\to\infty$ is evidently erroneous.

On the other hand, the uniform electron gas with a small number of particles can be studied in detail using the exact diagonalization technique, as has been recently done in few-electron parabolic quantum dots \cite{Mikhailov02b,Mikhailov02e,Merkt91,Wagner92,Mikhailov02c,Tavernier03}. One can expect, for example, the interaction induced Fermi liquid -- Wigner solid crossover, similar to that predicted in quantum dots \cite{Egger99a,Mikhailov02b}. Such studies could open up new interesting directions of research and lead to a better understanding of the physics of Coulomb interaction in solids.


\begin{thebibliography}{18}
\bibitem[{Gel()}]{GellMann}
\bibinfo{note}{In the second and higher orders of the non-degenerate
  perturbation theory one faced the problem of divergent Coulomb integrals, see
  \cite{GellMann57} and discussion in \cite{Mahan90,Kittel87,Fulde95}. In the
  degenerate perturbation theory such a problem does not emerge since one gets
  the higher accuracy by increasing the size of the Hamiltonian matrix and all
  the Coulomb matrix elements $\langle \Psi_i|V|\Psi_j\rangle$ are finite (of
  order unity in Rydberg units).}
\end{thebibliography}

\end{document}